\documentclass[journal=jceaax,manuscript=article]{achemso}

\usepackage[version=3]{mhchem} % Formula subscripts using \ce{}
\usepackage{comment}
\usepackage{longtable}
\usepackage{makecell}
\usepackage{tablefootnote}
\SectionNumbersOn
\author{Tae Jun Yoon}
\email{tyoon@lanl.gov}
\author{Erica P. Craddock}
\author{Katie A. Maerzke}
\author{Robert P. Currier}
\author{Alp T. Findikoglu}
\affiliation{Los Alamos National Laboratory, Los Alamos, NM 87545, United States}

\title{Electrical Conductances and association constants in dilute aqueous NdCl\textsubscript{3} solutions from 298 to 523 K along an isobar of 25 MPa}

\begin{document}

\begin{tocentry}
	\begin{center}
		\includegraphics{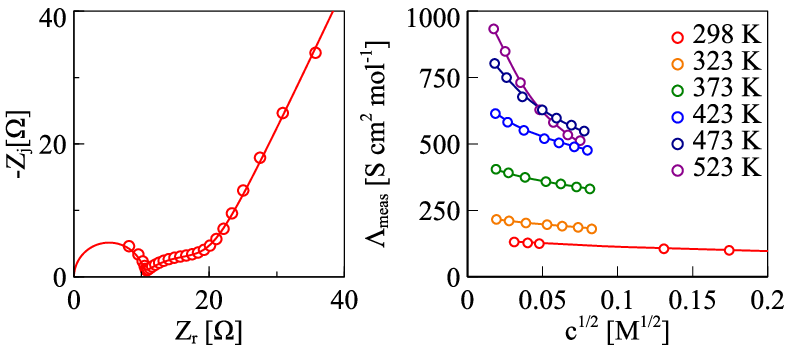}
	\end{center}
\end{tocentry}

\begin{abstract}
	Electrical measurements were performed in dilute aqueous NdCl\textsubscript{3} solutions from 298 to 523 K along the 25 MPa isobar to obtain limiting conductances and association constants. The specific conductance data were estimated using a continuous flow cell and a Markov Chain Monte Carlo (MCMC) correction algorithm. The limiting conductances for the salts in water were derived by regressing the mean spherical approximation (MSA) conductance model and speciation analyses based on the MCMC algorithm and the Deep Earth Water (DEW) model. The limiting conductances derived from the experimental data agree well with a predictive correlation proposed by Smolyakov, Anderko, and Lencka. Only the first association constant between neodymium and chloride could be derived at low temperatures ($< 373$ K) due to the apparent large statistical uncertainty of the second association constant. Above 373 K, both association constants could be derived and show a reasonable agreement with Migdisov and Williams-Jones and Gammons et al.
\end{abstract}

\pagebreak
\section{Introduction}
Rare earth elements (REEs), a group of elements comprising yttrium, scandium, and lanthanides, are of great importance from an industrial point of view. Their unique physicochemical properties make them essential materials for a variety of state-of-the-art technologies. Neodymium, for instance, is utilized for yttrium aluminum garnet in laser and high-strength permanent magnets. Since the high-strength magnets are used in high-tech electronic and energy technologies, a stable supply chain for these critical materials is necessary for transferring to a carbon-neutral world. 

Unfortunately, it is challenging to retrieve REEs in an environmentally benign fashion. Current commercial processes to obtain and refine one ton of REEs discharge 60,000 m\textsuperscript{3} of waste gas, 2,000 m\textsuperscript{3} of sewage water, and $1-1.4$ tons of radioactive waste.\cite{liu2016REEs} Commercial liquid-liquid extraction processes also require up to hundreds of stages of mixers and settlers to separate REEs.\cite{xie2014critical} In order to produce REEs in a more environmentally benign manner, various alternative REE resources and techniques are being explored.

Understanding the thermodynamic and kinetic behavior of REEs in natural and artificial systems is required to develop such environmentally-friendly processes.\cite{das2017rare,das2019rare,van2009molecular,finney2019ion,schijf2021speciation,cetiner2005aqueous,haas1995rare,mioduski2009iupac} A myriad of theoretical and experimental methods have been adopted to study the thermophysical behavior of aqueous REE solutions. According to previous literature, REEs in an aqueous environment show complicated physicochemical characteristics. For instance, REE chlorides do not form ion aggregates in ambient water, even in modestly concentrated solutions, despite the high valence number (+3) of the REE cations. Bernard et al. \cite{bernard1992conductance} demonstrated that the equivalent conductances in lanthanum chloride (LaCl\textsubscript{3}) solution at room temperature calculated without considering ion pair formation agree with the experimental data. Recent molecular simulations and spectroscopic studies reveal that this non-associating behavior arises from a strong hydration shell.\cite{finney2019ion,yoon2020dielectric} On the other hand, REE sulfates are highly aggregated in an aqueous environment,\cite{migdisov2006spectrophotometric,spedding1954conductances} but their solubilities in water show a retrograde dependence on temperature. 

Electrical measurements and analyses in aqueous REE solutions have also been conducted\cite{spedding1954conductances,spedding1952conductances,spedding1974electrical,apelblat2011representation} but mainly focused on the solutions at ambient conditions. A recent compilation of literature data by Corti\cite{corti2008electrical} indicates that there is only one study about the experimental measurements of specific conductance in NdCl\textsubscript{3} (\textit{aq}) at elevated temperatures.\cite{ismail2003conductivity} Ismail et al. measured the specific conductance in dilute rare earth chloride solutions (0.001 mol/kg) from 373 to 673 K along different isobars. They reported that NdCl\textsubscript{3} (\textit{aq}) showed the smallest electrical conductivity among four rare earth chloride solutions and had a conductance maximum at approximately 548 K.

REE solutions at elevated temperatures have been mainly studied based on other theoretical/empirical methods. Using experimental data at 298 K and 1 bar, Haas et al.\cite{haas1995rare} proposed thermodynamic parameters for predicting thermodynamic properties of rare earth solutions from 1 to 5,000 bars, and 273 to 1,273 K based on the Deep Earth Water (DEW) model (also known as the extended Helgeson-Kirkham-Flowers model).\cite{helgeson1981theoretical,tanger1988calculation,shock1989calculation,sverjensky2014water,huang2019extended,chan2021dewpython} Gammons et al.\cite{gammons1996aqueous} derived the association constants from solubility experiments and reported that a higher degree of association occurs at high temperatures, compared to the theoretical results proposed by Haas et al.\cite{haas1995rare} and Wood et al.\cite{wood1990aqueous} Migdisov and Williams-Jones\cite{migdisov2002spectrophotometric} conducted an ultraviolet spectroscopic study to estimate the association constants. Their results suggest that the first and the second association constants between neodymium and chloride should be lower than Haas et al. at low temperatures, while they become higher at elevated temperatures.

This work aims to obtain the limiting conductance and association constants in dilute aqueous neodymium chloride [NdCl\textsubscript{3} (\textit{aq})] solutions at elevated temperatures and pressures by measuring and analyzing electrical conductances. A continuous flow cell equipped with an \textit{in-situ} conductometric sensor was utilized to obtain their equivalent conductances at 25 MPa. The conductance data were analyzed based on the Metropolis-Hastings algorithm together with mean spherical approximation (MSA) and chemical speciation analyses proposed in our earlier works.\cite{yoon2021pyoecp,yoon2021situ,yoon2021selective}

\section{Methods}
\subsection{Materials}
Distilled water (H\textsubscript{2}O, Kroger\textsuperscript{\tiny\textregistered}) was purchased in a local store. The specific conductivity of the distilled water was determined as $<10^{-6}\ \mathrm{S/cm}$ by a handheld conductometer (EW-19601-03, Cole-Parmer\textsuperscript{\tiny\textregistered}). Neodymium chloride hexahydrate ($\mathrm{NdCl\cdot(H_2O)_6}$, $>99.0 \%$, Sigma Aldrich) was purchased from Sigma Aldrich. No additional purification was performed on the substances.
\subsection{Continuous Flow Cell}
\begin{figure}
	\includegraphics[width=\textwidth]{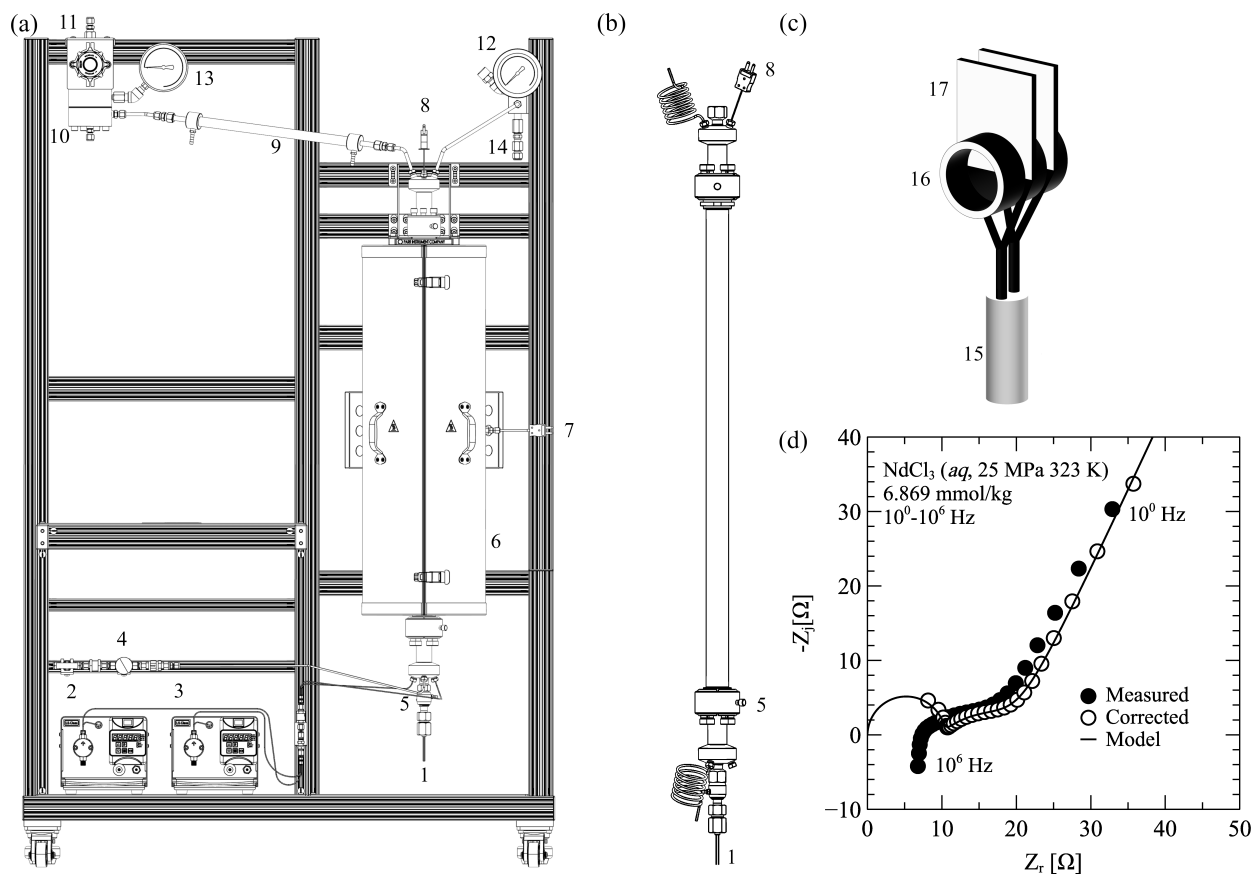}
	\caption{Schematic diagrams of (a) continuous flow unit and (b) vertical flow cell equipped with (c) a pair of in-situ electrodes: 1, electrical feedthrough; 2 and 3, LS-class HPLC pumps; 4, purge line; 5, chiller; 6, split-tube type heater; 7, outside thermocouple; 8, inside thermocouple; 9, single-pass heat exchanger; 10, back pressure regulator; 11, forward pressure regulator (dome); 12, vessel pressure indicator; 13, dome pressure indicator; 14, pressure relief device; 15, two-bore alumina rod; 16, alumina support; 17, platinized platinum plates. Since instrumental artifacts (e.g., wire inductance and parasitic capacitance) influence the impedance spectra, the correction procedure proposed in our earlier works was used to estimate the bulk resistance $Z_\mathrm{r}$ and bulk reactance $-Z_\mathrm{j}$ in a frequency range from $10^{0}$ to $10^{6}$ Hz as shown in (d).}
	\label{Fig1:flow-cell-configuration}
\end{figure}
Feed solutions were prepared using an electronic scale (Model MS 204S/03, Mettler Toledo\textsuperscript{\tiny\textregistered}). Figure \ref{Fig1:flow-cell-configuration} shows the schematic of the continuous flow unit. The instrumental details can be found elsewhere.\cite{yoon2021situ} Here, only a brief description of the measurement protocol is given. Two high-pressure pumps were used to supply brine solutions from the bottom into a tubular cell. Two thermocouples were used to measure and control the cell temperature. One thermocouple was inserted in the tubular cell to measure the solution temperature directly, while the other measured the cell surface temperature. A split-type heater was used to increase the temperature to the target temperatures. The operating pressure was regulated by a dome-type back pressure regulator connected to a 40 MPa (6,000 psi) argon cylinder. The precision of the pressure transducer installed in this unit was 0.007 MPa ($1\ \text{psi}$). A pair of platinized platinum electrodes were inserted into the cell and connected to a frequency response analyzer (Solartron 1260A, Ameteck, Inc.). A pair of titanium wires (Ti-6Al-4V, Nexmetal Corporation), which were oxidized at 873 K in air,\cite{velten2002preparation} were welded to the electrodes and enveloped in a two-bore alumina rod to immobilize the electrode. The cell was chemically passivated using a citric acid solution to prevent excessive corrosion.\cite{yasensky2009citric}
\subsection{Specific Conductance Estimation}
We followed the specific conductance estimation protocol suggested in our earlier work for sodium chloride solutions.\cite{yoon2021situ} Before every measurement, the cell was flushed with distilled water at elevated temperatures and pressure. After the cell was cooled to 298 K, it was cleaned by supplying additional distilled water at ambient conditions. The cell was dried at 393 K overnight. The \textit{Open} circuit measurement was done before introducing the prepared solutions. The cell was first flushed with one liter of the sample solution to remove any remaining water or corrosion products. Then, it was pressurized to 25 MPa at 298 K and heated to the target temperature(s). 

To ensure that (1) no neodymium salt precipitates and (2) no considerable amount of impurities were generated from the cell parts at the target temperatures, an ultraviolet spectrophotometer (Evolution 201, Thermo Scientific, Inc.) was used to analyze the effluent from the continuous flow unit. The ultraviolet calibration curve was constructed by measuring the absorbance of standard neodymium solutions around 574 nm. For the calibration curve, see the Supporting Information in our earlier work.\cite{yoon2021selective} The specific conductance and the neodymium concentration in the effluent did not show a significant change ($<2\ \%$) until the solution temperature reached 523 K. Thus, the \textit{in-situ} specific conductance measurements were performed below 523 K.

When the system temperature read by the inner thermocouple matches the target temperature, the solution impedance ($Z=Z_\mathrm{r}+iZ_\mathrm{j}$) was measured from $10^0$ to $10^6$ Hz in triplicate. Here, $Z_\mathrm{r}$ and $Z_\mathrm{j}$ are real and imaginary parts of the complex impedance, and $i$ is the imaginary unit number $i^2=-1$. The measured impedance spectra $Z_\mathrm{m}$ were averaged and corrected by an \textit{Open/Short} correction algorithm to estimate the impedance of the material under test (MUT). According to the algorithm, the averaged impedance spectrum is given as:
\begin{equation}
	Z_\mathrm{m}(f)=Z_\mathrm{O}(f)+\frac{1}{Z_\mathrm{O}^{-1}(f)+Z_\mathrm{MUT}^{-1}(f)}
\end{equation}
where $Z_\mathrm{O}$ $Z_\mathrm{S}$, $Z_\mathrm{MUT}$ are \textit{Open}, \textit{Short}, and MUT impedances. By fitting an equivalent circuit model,\cite{yoon2021situ} the bulk resistance in a solution ($R_\mathrm{MUT}$) was estimated (see Figure \ref{Fig1:flow-cell-configuration} (d) and our earlier work\cite{yoon2021situ} for the detailed description). The bulk solution resistance was converted to the specific conductance ($\kappa$) as:
\begin{equation}
	\kappa=\frac{1}{R_\mathrm{MUT}}\frac{l}{A}
	\label{eq: specific-conductance}
\end{equation}
where $l/A$ is the cell constant. The cell constant was calculated as 2.985$\pm$0.050 $\mathrm{m^{-1}}$ at room temperature using the specific conductance data provided by Spedding\cite{spedding1952conductances} and did not show a noticeable degradation.
\subsection{Data Analysis}
The specific conductance data estimated from Eq. \ref{eq: specific-conductance} was converted to equivalent conductance $\Lambda$ ($\mathrm{S\ {cm}^2/mol}$). It is defined as
\begin{equation}
	\Lambda=\frac{\kappa}{N}=\frac{\kappa}{\sum_i^\mathrm{cations}c_{i}z_{i}}=\frac{\kappa}{\sum_j^\mathrm{anions}c_{j}|z_{j}|}
\end{equation}
where $N$ denotes the equivalent concentration (normality, eq/L), and $c_{k}$ and $z_{k}$ are the molar concentration and the charge of an ion $k$. In converting the concentration unit from molality to molarity, it was assumed that the solution density is equal to that of pure water at the same thermodynamic conditions, considering that all solutions are very dilute.\cite{gruszkiewicz1997conductance,quist1968electrical}

After the pioneering works by Debye and H\"uckel\cite{debye1923theory} and Onsager,\cite{onsager1926theorie} a variety of conductance models, including the Shedlovsky model,\cite{shedlovsky1932equation} Fuoss-Onsager model,\cite{fuoss1978review} Fuoss-Hsia-Fern\'andez-Prini model,\cite{fernandez1969conductance} Lee-Wheaton model,\cite{lee1978conductance1,lee1978conductance2,lee1979conductance3} Quint-Viallard model,\cite{quint1978electrical,quint1978electrophoretic,quint1978relaxation} Barthel model (low-concentration chemical model, lcCM),\cite{barthel1998physical} and mean spherical approximation (MSA) model,\cite{bernard1992conductance,turq1995conductance} were proposed. In the field of high-temperature and high-pressure aqueous chemistry, the  Turq-Blum-Bernard-Kunz (TBBK) model \cite{turq1995conductance} and the Fuoss-Hsia-Fern\`andez-Prini (FHFP) model\cite{fernandez1969conductance} have been widely used. Since the FHFP model is suitable only for symmetric electrolytes, the TBBK model was adopted in this work. 

The TBBK model calculates the equivalent conductance based on the MSA theory. The equivalent conductance contribution of a single species $k$ ($\lambda_k$) is given as:
\begin{equation}
	\lambda_k=\lambda_k^\infty\left(1+\frac{\delta v_k}{v_k}\right)\left(1+\frac{\delta X}{X}\right)
\end{equation}
where $\lambda_k^\infty$ is the limiting conductance of the species $k$, $\delta v_k/v_k$ is the electrophoretic contribution, and $\delta X/X$ is the relaxation contribution, including the hydrodynamic term. Each term is a complex function of the size parameter $\sigma_k$, concentration $c_k$, static permittivity $\epsilon_\mathrm{r}$, and solvent viscosity $\eta$. See Supporting Information for all expressions (all known misprints\cite{bianchi2000comparison,sharygin2001tests} in the original article by Turq et al.\cite{turq1995conductance} were corrected). Static permittivity and viscosity of pure water were calculated from Fern\'andez et al.\cite{fernandez1997formulation} and Huber et al.\cite{huber2009new} The solution densities were assumed to be equal to the solvent densities calculated from the Wagner-Pru\ss\ equation of state\cite{wagner2002iapws}. 

The limiting conductance $\lambda_{k}^\infty$ of the species $k$ is defined as
\begin{equation}
	\lambda_{k}^\infty=\lim_{c_{{k}}\rightarrow0}\lambda_{k}(c_{k})
\end{equation}
where $c_{k}$ is the molar concentration of the species $k$. It is directly related to the limiting diffusion coefficient of the species $k$ ($D_k^\infty$) as:\cite{turq1995conductance,bernard1992self,dufreche2002ionic}
\begin{equation}
	\frac{D_k^\infty}{\lambda_k^\infty}=\frac{RT}{z_kF^2}
\end{equation}
where $R$ is the universal gas constant, $z_k$ is the charge, and $F$ is the Faraday constant.

Although the TBBK model has been widely utilized, several inconsistencies were reported between the model predictions and the empirical results. Thus, the original TBBK model was modified following the suggestions in earlier studies.\cite{bianchi2000comparison,sharygin2001tests,reilly1969prediction,wu1988cross,anderko1997computation} First, the Bjerrum distance between cation and anion pairs was used as the size parameter for calculating the electrostatic part of the activity coefficient $y_i$ in the MSA formalism.\cite{sharygin2001tests} Second, the first-order mixing rule proposed by several authors\cite{reilly1969prediction,miller1996binary,wu1988cross,anderko1997computation} and tested by Sharygin et al.\cite{sharygin2001tests} was adopted to calculate the equivalent conductance, considering ion-pair formation in the system. This procedure was essential since neodymium ion is poly-valent and forms hydroxide complexes $\mathrm{Nd(OH)}_{n}$ shifting the acid-base equilibrium in the system.\cite{haas1995rare} The mixing rule is given as:
\begin{equation}
	\Lambda=\sum_{i}^{\mathrm{cations}}\sum_{j}^{\mathrm{anions}}f_if_jl_{ij}(\Gamma)
\end{equation}
where the sum is over all cation-anion pairs $(i,j)$ in the system. The normalized fraction of a species $k$ ($f_k$) is defined as $f_k\equiv{c_k}|z_k|/N$, and the pair equivalent conductance contribution $l_{ij}(\Gamma)$ is calculated based on the original TBBK equations given in the Supporting Information. Note that the screening parameter $\Gamma$ was calculated by considering not just a single pair $(i,j)$, but all species in the system.\cite{sharygin2006dissociation} 

The ion pairs and free ions concentrations were calculated using the ion speciation algorithm developed in our earlier work.\cite{yoon2021selective} In this algorithm, chemical equilibria between different species should be first determined. In NdCl\textsubscript{3} solutions, the following equilibria can be considered.
\begin{subequations}
	\begin{equation}
	\mathrm{Nd^{3+}+\mathit{n}Cl^{-}\rightleftharpoons (NdCl_\mathit{n})^{3-\mathit{n}}}\ (n=1,\ 2,\ 3,\ \text{and}\ 4)
	\end{equation}
	\begin{equation}
	\mathrm{Nd^{3+}+\mathit{n}OH^{-}\rightleftharpoons (Nd(OH)_\mathit{n})^{3-\mathit{n}}}\ (n=1,\ 2,\ 3,\ \text{and}\ 4)
	\end{equation}
	\begin{equation}
	\mathrm{H^++OH^-\rightleftharpoons{H_2O}}
	\end{equation}
\end{subequations}

We neglected the third and fourth chemical equilibria between $\mathrm{Nd}^{3+}$ and $\mathrm{Cl}^{-}$, considering their population is much lower than the other species; the association constants cannot be estimated reliably. Unlike neutral salts like sodium chloride, the neodymium ion forms a pair with hydroxide ion(s) $\mathrm{Nd(OH)}_n$. The formation of neodymium hydroxide complexes increases the hydrogen ion concentration, which could significantly contribute to the total equivalent conductance. Thus, the neodymium-hydroxide equilibria were always included. The Bandura-Lvov equation was used to calculate the self-ionization constant.\cite{bandura2006ionization} The activity coefficients $y_\mathrm{i}$'s were calculated by using the modified protocol suggested by Sharygin et al.\cite{sharygin2001tests} (see the Supporting Information for the detailed calculation procedure).
 
After calculating the association constants and activity coefficients, the whole set of equations, including the mass and charge balance (electroneutrality) equations, were solved simultaneously using the modified Powell method. 

\begin{figure}
	\includegraphics[width=0.8\textwidth]{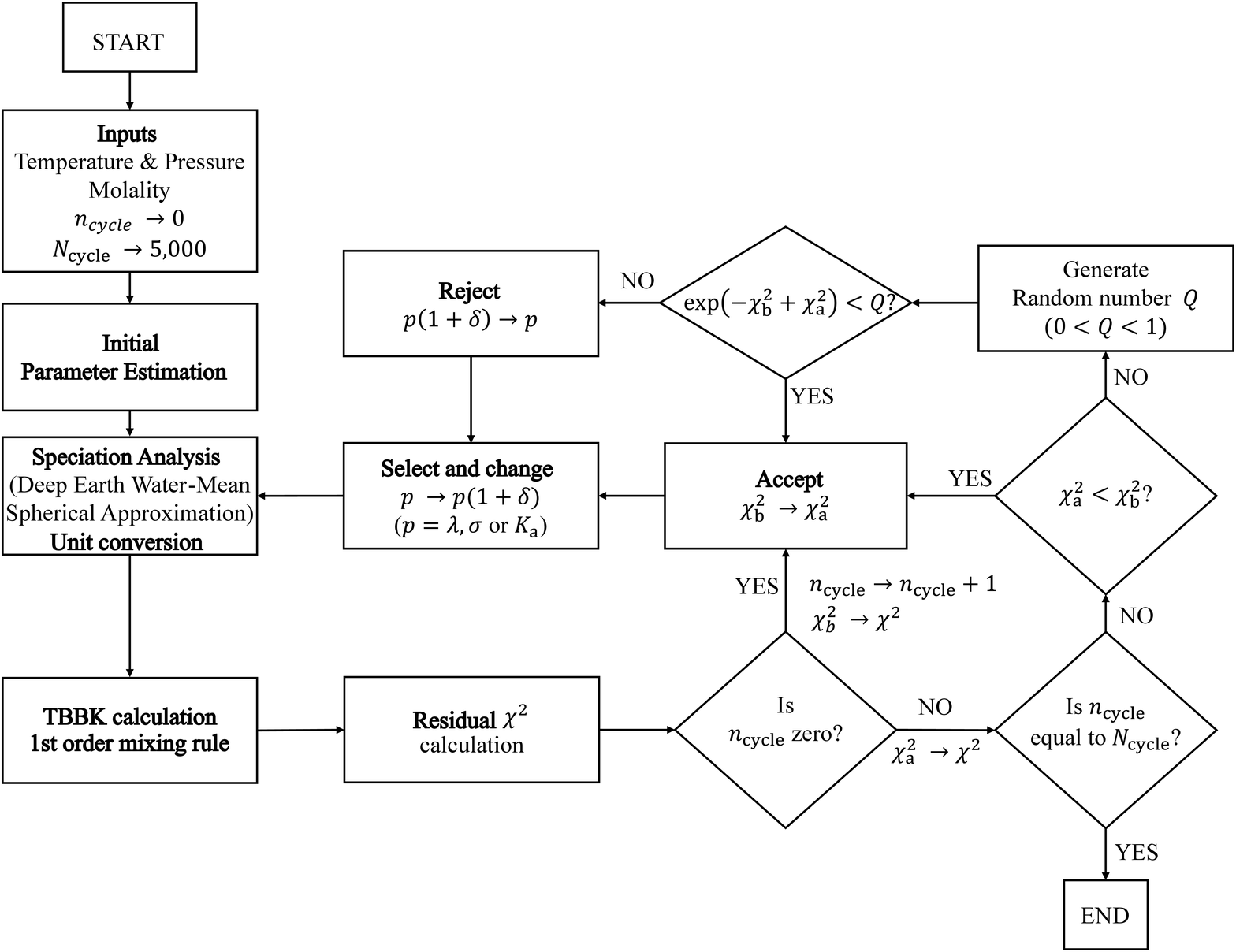}
	\caption{A model regression algorithm based on the Metropolis-Hastings (Markov Chain Monte Carlo) algorithm. For calculation details (parameter initialization, Turq-Blum-Bernard-Kunz equation calculation, and first order mixing rule), see the main text and the Supporting Information.}
	\label{Fig2:algorithm}
\end{figure}
Assembling all addressed points, an iterative procedure was designed to regress the TBBK equation to the experimental data (Figure \ref{Fig2:algorithm}), which is similar to that proposed by Zimmerman,\cite{zimmerman2012limiting2} except that the solution density is assumed to be equal to the solvent density and the Metropolis-Hastings algorithm is used for model regression. In this procedure, the limiting conductance $\lambda_k^\infty$ and size parameters are first estimated. The limiting conductance of $\mathrm{Cl^\mathrm{-}}$ is estimated using an empirical correlation recently proposed by Zimmerman, Arcis, and Tremaine.\cite{zimmerman2012limiting} It is given as:
\begin{equation}
	\lambda_\mathrm{Cl}^\infty={A_1}\eta^{A_2+A_3/\rho_\mathrm{w}}
	\label{eq:ZAT-correlation}
\end{equation}
where $A_1$, $A_2$, and $A_3$ are adjustable parameters ($A_1=1.137\pm0.006$, $A_2=-0.925\pm0.002$, and $A_3=34.44\pm1.24$), $\eta$ is the solvent viscosity in Poise, and $\rho_\mathrm{w}$ is the solvent density in kg/m\textsuperscript{3}. The limiting conductivities of hydrogen and hydroxide ions ($\lambda_\mathrm{H}$ and $\lambda_\mathrm{OH}$) were calculated from an empirical correlation proposed by Marshall.\cite{marshall1987reduced} The limiting conductivity of $\mathrm{Nd}^{3+}$ ($\lambda_\mathrm{Nd}^\infty$) was assigned arbitrarily.

The crystallographic radii taken from Marcus\cite{marcus1988ionic} were mainly used to fit the experimental data except those at 298 K (see Sec. \ref{sec:Results-model-validation}). The first and the second association constants between Nd and Cl ions ($K_\mathrm{NdCl}$ and $K_\mathrm{NdCl_2}$) were initially estimated using the DEWPython module.\cite{chan2021dewpython} The DEW parameters provided by Migdisov and Williams-Jones\cite{migdisov2002spectrophotometric} and those by Haas et al.\cite{haas1995rare} were used as initial estimates of $K_\mathrm{NdCl}$ and $K_\mathrm{NdCl_{2}}$. The converged values with different initial association constants were not significantly different. The other association constants for neodymium and hydroxide ions were calculated based on the DEW parameters proposed by Haas et al.\cite{haas1995rare} and were not altered in the fitting procedure. 

The limiting conductance and the size parameter of an ion pair P consisting of the $m$ number of ions were estimated as:
\begin{subequations}
	\begin{equation}
	\lambda_\mathrm{P}^\infty=\frac{|z_\mathrm{P}|}{\left(\sum_{i=1}^{m}(z_i/\lambda_i^\infty)^3\right)^{1/3}}
	\end{equation}
	\begin{equation}
	\sigma_\mathrm{P}=\left(\sum_{i=1}^{m}\sigma_i^3\right)^{1/3}
	\end{equation}
\end{subequations}

After the initialization, the limiting conductivity of the neodymium ion $\lambda_\mathrm{Nd}^\infty$ and the association constants $K_\mathrm{NdCl}$ and $K_\mathrm{NdCl_{2}}$ were changed based on the Metropolis-Hastings algorithm.\cite{yoon2021pyoecp} In this algorithm, one of the adjustable parameters $p$ is randomly chosen and changed to $p'=p(1+\delta)$, where $\delta$ is a random number between $-M$ and $M$. The magnitude $M$ was 0.05 in this work. When $N_\mathrm{p}$ data points are given, the residual between the model calculation and the experimental data is evaluated as:
\begin{equation}
	\chi^2=\sum_{i=1}^{N_{P}}\left(\Lambda_i^\mathrm{calc}-\Lambda_i^\mathrm{meas}\right)^2
\end{equation}
where $\Lambda_i^\mathrm{calc}$ and $\Lambda_i^\mathrm{meas}$ are the calculation results and experimental data point at the $i^\mathrm{th}$ point. When the residual calculated from the changed parameters is lower than that from the original parameters, the changed parameter $p'$ is accepted. If not, a random number between zero and one is generated, and compared to $\exp(-\chi_\mathrm{after}^2+\chi_\mathrm{before}^2)$. If the random number is smaller than $\exp(-\chi_\mathrm{after}^2+\chi_\mathrm{before}^2)$, the result is accepted. Otherwise, the result is rejected. This procedure is repeated until all parameters are converged. In fitting the model to the equivalent conductance data, the second association constant $K_\mathrm{NdCl_{2}}$ did not show a stable convergence at low temperatures. It showed a variation from $10^{-7}$ to $10$. This result suggests that $K_\mathrm{NdCl_{2}}$ has no significant contribution to the conductance data. (see Sec. \ref{sec:Results-Discussion}) In this case, only the first association step was considered in the fitting procedure.

After the fitting procedure, the discrepancy between the model and the experimental data was evaluated using absolute average relative deviation (AAD). It is defined as:
\begin{equation}
	\mathrm{AAD}=\left(\frac{|\Lambda^\mathrm{calc}-\Lambda^\mathrm{meas}|}{\Lambda^\mathrm{meas}}\right)100\ [\%]
\end{equation}
\section{Results and discussion}
\label{sec:Results-Discussion}
\subsection{Model validation}
\label{sec:Results-model-validation}
\begin{figure}
	\includegraphics{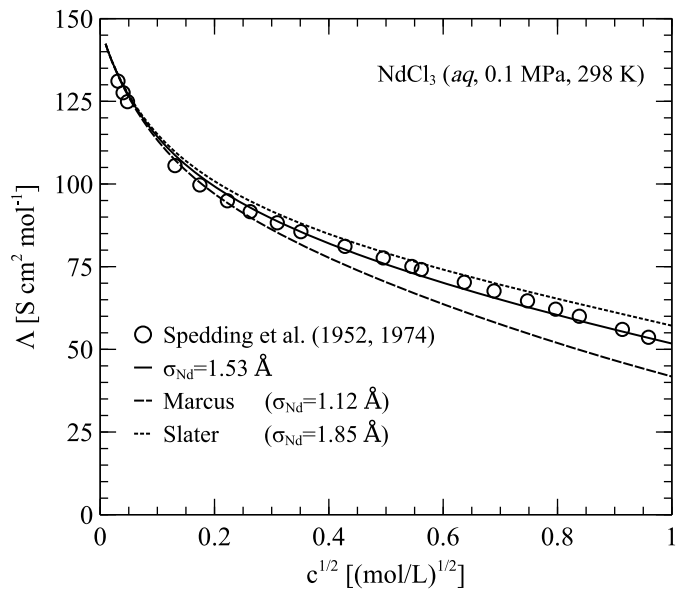}
	\caption{TBBK model fit to the equivalent conductance data in NdCl\textsubscript{3} (\textit{aq}) measured by Spedding and coworkers at 298 K and 1 bar. The fitting result was not significantly influenced by the ion pair formation between chloride and neodymium. Rather, the goodness of fit was affected by the size parameter of the neodymium ion.}
	\label{Fig3:literature-data}
\end{figure}
Before analyzing the experimental data obtained in this work, we attempted to fit the TBBK model to literature data taken from Spedding and coworkers.\cite{spedding1952conductances,spedding1974electrical} (Figure \ref{Fig3:literature-data}) The fitting concentration range was limited to $c\leq1$ mol/L, considering the model limitation proposed by Turq et al.\cite{turq1995conductance} The model calculation results did not change significantly when the ion pair formation between neodymium and chloride ions was neglected. Instead, the deviation between the model and the empirical data depended on the size parameter $\sigma_\mathrm{Nd}$. When the crystallographic radius taken from Marcus was used ($\sigma_\mathrm{Nd}=1.12$~\AA), the predicted conductance was close to the experimental data in dilute solutions ($c^{1/2}<0.25 \mathrm{(mol/L)}^{1/2}$, AAD 1.246 \%) but became lower than the experimental data in concentrated solutions (overall AAD 8.29 \%). When the Slater radius\cite{slater1964atomic} ($\sigma_\mathrm{Nd}=1.85$ \AA) was used, the model over-predicted the conductance within the fitting concentration range (AAD 3.85 \%). By fixing the limiting conductance of the neodymium ion to be 69.8 S cm\textsuperscript{2}/mol,\cite{apelblat2011representation,spedding1974electrical} the optimal ionic radius to fit the experimental data within $c<1.0\ \mathrm{mol/L}$ was obtained as 1.53 \AA~with an overall AAD of 1.93 \%. When only dilute solutions were fitted, the optimal radius became close to the crystallographic radius $\sigma_\mathrm{Nd}=1.12$~\AA. The dependence of the optimal radius on the ionic strength of the system may indicate that a rigid hydration shell is formed in NdCl\textsubscript{3} (\textit{aq})\cite{finney2019ion,yoon2020dielectric}. Since this work only focuses on dilute solutions at elevated temperatures, we used the crystallographic radius throughout.\cite{marcus1988ionic}

It is noteworthy that the above results are not sensitive to the association constants. That is, the calculation results do not vary significantly depending on the association constants. Moreover, the results show a similar behavior without considering the association. The insensitivity of the conductance data to small association constants does not arise from the deficiency in the TBBK model. Indeed, rare earth chlorides in water at room temperature are sometimes regarded as non-associating solutes. For instance, Apelblat \cite{apelblat2011representation} used the Quint-Viallard model\cite{quint1978electrical,quint1978electrophoretic,quint1978relaxation} for dilute NdCl\textsubscript{3} solutions at 298 K without considering the ion association. Bernard et al.\cite{bernard1992conductance} demonstrate that the non-associative model can predict the equivalent conductances in lanthanum chloride solutions up to 1 mol/L. Thus, the insensitivity of the conductance data to association constants may indicate the limit of the conductance method to calculate the association constants in weakly associating electrolyte solutions.\cite{bianchi2000comparison} The statistical uncertainty of the second association constants was also large at low temperatures studied in this work. (Sec. \ref{sec:Results-main})

\subsection{Model regression}
\label{sec:Results-main}
\begin{longtable}{@{\extracolsep{\fill}}cccccc@{}}
	\caption{Specific conductance ($\kappa_\mathrm{meas}$) and equivalent conductance ($\lambda_\mathrm{meas}$) data and the TBBK model regression results ($\lambda_\mathrm{calc}$) at different temperatures and concentrations along an isobar of 25 MPa. The unit eq/L indicates equivalent molarity.}
	\tabularnewline
	\hline
	{m$\times10^{3}$} & {$N\times10^{3}$} & $\kappa_\mathrm{meas}$ & $\lambda_\mathrm{meas}$ & $\lambda_\mathrm{calc}$ & {AAD}\\
	$\mathrm{[mol/kg]}$ & $\mathrm{[eq/L]}$ & $\mathrm{[{\mu}S\ cm^{-1}]}$ & $\mathrm{[S\ cm^{2}\ mol^{-1}]}$ & $\mathrm{[S\ cm^{2}\ mol^{-1}]}$ & $\mathrm{[\%]}$\\
	\hline \endfirsthead
	\hline
	m$\times10^{3}$ & $N\times10^{3}$ & $\kappa_\mathrm{meas}$ & $\lambda_\mathrm{meas}$ & $\lambda_\mathrm{calc}$ & AAD\\
	$\mathrm{[mol/kg]}$ & $\mathrm{[eq/L]}$ & $\mathrm{[{\mu}S\ cm^{-1}]}$ & $\mathrm{[S\ cm^{2}\ mol^{-1}]}$ & $\mathrm{[S\ cm^{2}\ mol^{-1}]}$ & $\mathrm{[\%]}$\\
	\hline \endhead
	\hline \multicolumn{6}{r}{Continued on next page}
	\endfoot		 
	\hline \endlastfoot
	\multicolumn{6}{c}{$T=323$ K, $p=25$ MPa, $\rho_\mathrm{w}=998.7$ kg/m\textsuperscript{3}}\\
	0.371	&	1.111	&$	240.1	$&$	216.1		\pm	1.6	$&	215.9	&	0.10	\\
	0.767	&	2.297	&$	483.1	$&$	210.3		\pm	2.3	$&	209.8	&	0.22	\\
	1.516	&	4.540	&$	919.8	$&$	202.6		\pm	0.9	$&	202.8	&	0.11	\\
	2.815	&	8.428	&$	1654.8	$&$	196.3		\pm	1.5	$&	195.2	&	0.61	\\
	3.989	&	11.939	&$	2277.1	$&$	190.7		\pm	0.3	$&	190.3	&	0.23	\\
	5.444	&	16.287	&$	3032.8	$&$	186.2		\pm	0.6	$&	185.6	&	0.32	\\
	6.869	&	20.543	&$	3702.7	$&$	180.2		\pm	0.4	$&	181.9	&	0.94	\\
	\multicolumn{6}{c}{$T = 348$ K, $p = 25$ MPa, $\rho_\mathrm{w} = 985.7$ kg/m\textsuperscript{3}}\\
	0.371	&	1.097	&$	333.5	$&$	304.1		\pm	1.1	$&	305.1	&	0.32	\\
	0.767	&	2.267	&$	669.6	$&$	296.3		\pm	0.6	$&	295.3	&	0.32	\\
	1.516	&	4.481	&$	1274.5	$&$	287.0		\pm	1.0	$&	284.4	&	0.91	\\
	2.815	&	8.318	&$	2269.5	$&$	275.6		\pm	0.7	$&	272.8	&	0.99	\\
	3.989	&	11.784	&$	3129.5	$&$	266.8		\pm	1.2	$&	265.6	&	0.44	\\
	5.444	&	16.075	&$	4158.7	$&$	259.4		\pm	0.9	$&	258.7	&	0.27	\\
	6.869	&	20.276	&$	5191.3	$&$	256.0		\pm	0.8	$&	253.3	&	1.06	\\
	\multicolumn{6}{c}{$T = 373$ K, $p = 25$ MPa, $\rho_\mathrm{w} = 969.7$ kg/m\textsuperscript{3}}\\
	0.371	&	1.079	&$	436.8	$&$	404.8		\pm	1.3	$&	404.2	&	0.15	\\
	0.767	&	2.231	&$	872.9	$&$	391.3		\pm	2.3	$&	389.8	&	0.38	\\
	1.516	&	4.409	&$	1650.2	$&$	374.3		\pm	1.3	$&	374.4	&	0.03	\\
	2.815	&	8.184	&$	2931.4	$&$	358.2		\pm	0.9	$&	358.5	&	0.08	\\
	3.989	&	11.594	&$	4052.1	$&$	349.5		\pm	0.8	$&	348.6	&	0.25	\\
	5.444	&	15.816	&$	5348.5	$&$	338.2		\pm	0.4	$&	339.4	&	0.36	\\
	6.869	&	19.950	&$	6598.1	$&$	330.7		\pm	1.2	$&	332.1	&	0.43	\\
	\multicolumn{6}{c}{$T = 398$ K, $p = 25.0$ MPa, $\rho_\mathrm{w} = 951.3$ kg/m3}\\
	0.371	&	1.058	&$	539.0	$&$	509.3		\pm	1.1	$&	508.7	&	0.12	\\
	0.767	&	2.188	&$	1063.1	$&$	485.8		\pm	2.9	$&	487.4	&	0.31	\\
	1.516	&	4.325	&$	2003.9	$&$	463.4		\pm	2.7	$&	465.2	&	0.41	\\
	2.815	&	8.028	&$	3537.4	$&$	440.6		\pm	3.7	$&	443.1	&	0.57	\\
	3.989	&	11.373	&$	4879.2	$&$	429.0		\pm	2.7	$&	429.8	&	0.19	\\
	5.444	&	15.515	&$	6407.3	$&$	413.0		\pm	2.1	$&	417.5	&	1.09	\\
	6.869	&	19.570	&$	7907.6	$&$	404.1		\pm	0.2	$&	408.0	&	0.96	\\
	\multicolumn{6}{c}{$T = 423$ K, $p = 25.0$ MPa, $\rho_\mathrm{w} = 930.4$ kg/m\textsuperscript{3}}\\
	0.371	&	1.035	&$	635.8	$&$	614.2		\pm	2.5	$&	614.8	&	0.09	\\
	0.767	&	2.140	&$	1245.1	$&$	581.7		\pm	1.6	$&	584.6	&	0.49	\\
	1.516	&	4.230	&$	2331.1	$&$	551.1		\pm	1.3	$&	553.4	&	0.42	\\
	2.815	&	7.852	&$	4081.8	$&$	519.8		\pm	1.2	$&	522.9	&	0.60	\\
	3.989	&	11.124	&$	5606.0	$&$	504.0		\pm	1.9	$&	505.0	&	0.21	\\
	5.444	&	15.176	&$	7419.7	$&$	488.9		\pm	3.4	$&	488.7	&	0.03	\\
	6.869	&	19.142	&$	9112.5	$&$	476.0		\pm	0.5	$&	476.4	&	0.07	\\
	\multicolumn{6}{c}{$T = 448$ K, $p = 25.0$ MPa, $\rho_\mathrm{w} = 907.2$ kg/m\textsuperscript{3}}\\
	0.371	&	1.009	&$	723.1	$&$	716.4		\pm	1.7	$&	715.0	&	0.19	\\
	0.767	&	2.087	&$	1399.7	$&$	670.7		\pm	1.1	$&	674.8	&	0.61	\\
	1.516	&	4.125	&$	2593.5	$&$	628.8		\pm	1.1	$&	631.4	&	0.41	\\
	2.815	&	7.657	&$	4496.7	$&$	587.3		\pm	3.6	$&	589.4	&	0.36	\\
	3.989	&	10.847	&$	6115.0	$&$	563.8		\pm	1.3	$&	565.3	&	0.28	\\
	5.444	&	14.798	&$	8028.2	$&$	542.5		\pm	3.3	$&	543.8	&	0.23	\\
	6.869	&	18.666	&$	9835.6	$&$	526.9		\pm	0.2	$&	527.6	&	0.13	\\	
	\multicolumn{6}{c}{$T = 473$ K, $p = 25.0$ MPa, $\rho_\mathrm{w} =  881.5$ kg/m\textsuperscript{3}}\\
	0.371	&	0.981	&$	787.5	$&$	802.9		\pm	3.4	$&	800.7	&	0.28	\\
	0.767	&	2.028	&$	1520.5	$&$	749.9		\pm	3.2	$&	748.2	&	0.22	\\
	1.516	&	4.008	&$	2714.3	$&$	677.3		\pm	4.6	$&	687.5	&	1.51	\\
	2.815	&	7.440	&$	4672.1	$&$	628.0		\pm	0.7	$&	629.2	&	0.20	\\
	3.989	&	10.540	&$	6295.2	$&$	597.3		\pm	1.1	$&	596.8	&	0.08	\\
	5.444	&	14.379	&$	8207.6	$&$	570.8		\pm	1.3	$&	568.5	&	0.41	\\
	6.869	&	18.138	&$	9949.0	$&$	548.5		\pm	1.4	$&	547.6	&	0.17	\\	
	\multicolumn{6}{c}{$T = 498$ K, $p = 25$ MPa, $\rho_\mathrm{w} = 853.0$ kg/m\textsuperscript{3}}\\
	0.371	&	0.949	&$	832.0	$&$	876.7		\pm	8.9	$&	869.5	&	0.83	\\
	0.767	&	1.962	&$	1579.8	$&$	805.2		\pm	2.5	$&	799.7	&	0.68	\\
	1.516	&	3.878	&$	2744.7	$&$	707.8		\pm	3.2	$&	714.8	&	1.00	\\
	2.815	&	7.199	&$	4671.9	$&$	649.0		\pm	1.5	$&	634.9	&	2.17	\\
	3.989	&	10.199	&$	6012.2	$&$	589.5		\pm	0.9	$&	592.1	&	0.44	\\
	5.444	&	13.914	&$	7769.7	$&$	558.4		\pm	2.8	$&	555.7	&	0.48	\\
	6.869	&	17.552	&$	9568.3	$&$	545.1		\pm	1.7	$&	529.6	&	2.86	\\	
	\multicolumn{6}{c}{$T = 523$ K, $p = 25$ MPa, $\rho_\mathrm{w}  = 821.1$ kg/m\textsuperscript{3}}\\
	0.371	&	0.914	&$	852.3	$&$	932.9		\pm	7.7	$&	930.1	&	0.31	\\
	0.767	&	1.889	&$	1602.3	$&$	848.3		\pm	1.7	$&	842.5	&	0.69	\\
	1.516	&	3.733	&$	2727.6	$&$	730.6		\pm	0.2	$&	732.1	&	0.20	\\
	2.815	&	6.931	&$	4355.9	$&$	628.5		\pm	2.6	$&	629.5	&	0.16	\\
	3.989	&	9.819	&$	5703.8	$&$	580.9		\pm	0.4	$&	575.9	&	0.86	\\
	5.444	&	13.396	&$	7151.3	$&$	533.8		\pm	1.5	$&	531.3	&	0.48	\\
	6.869	&	16.898	&$	8658.7	$&$	512.4		\pm	0.6	$&	499.6	&	2.49	
	\label{tab1:measured-conductance}
\end{longtable}
\begin{figure}
	\includegraphics{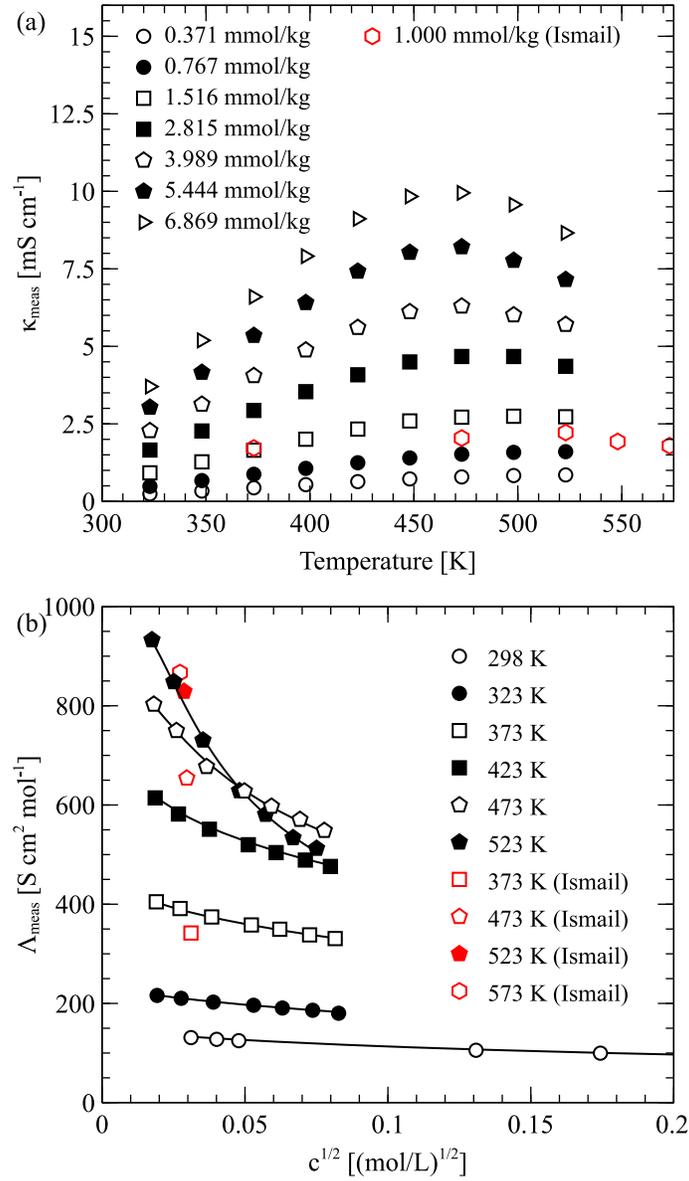}
	\caption{(a) Specific conductance estimated from the impedance spectra. (b) Equivalent conductance data (symbols) and the TBBK fitting results (lines). In (b), the data were represented with a 50 K interval for visual clarity (except the literature data by Spedding et al.\cite{spedding1952conductances,spedding1974electrical}) Red symbols were calculated from the experimental data measured by Ismail et al. (24.52 MPa, $m=1.00$ mmol/kg)\cite{ismail2003conductivity} They show a reasonable agreement with our data ($\text{AAD}<10 \%$).}
	\label{Fig4:main-r esult}
\end{figure}
Figure \ref{Fig4:main-result} (a) shows the specific conductance data estimated from the measured impedance spectra. All specific conductance data show an almost linear increase at $T<425$ K due to the increase in kinetic energy of individual ions and the increase in the hydrogen ion concentration. Above 425 K, the specific conductance curves become concave downward. At concentrations below 1.516 mmol/kg, no conductance maximum was observed. The specific conductance maximum is observed as the concentration increases and shifts to a lower temperature ($\sim473$ K). This result would arise from ion pair formation, as demonstrated in our earlier simulation work on NaCl.\cite{yoon2019electrical} Considering that (1) 1 mmol/kg NdCl\textsubscript{3} (\textit{aq}) has a conductance maximum at $\sim$548 K\cite{ismail2003conductivity} and (2) the specific conductance data obtained by Ismail et al.\cite{ismail2003conductivity} shows a reasonable agreement with ours, the measured specific conductance data seem reliable.

Figure \ref{Fig4:main-result} (b) shows the equivalent conductance data and the TBBK model regression results. The equivalent conductance data from 323 to 523 K are presented with a 50 K interval for visual clarity (see Table \ref{tab1:measured-conductance} for the equivalent conductance data estimated at other temperatures). The average AAD between the model and the experimental data was 0.5 \%. At low temperatures, the equivalent conductance shows a linear dependence on the square root of the concentrations ($c^{1/2}$), following the limiting law (Debye-H\"uckel-Onsager equation). As the temperature increases, the equivalent conductance curve along an isotherm becomes concave upwards, implying ion pair formation. The equivalent conductance data at 24.52 MPa measured by Ismail et al.\cite{ismail2003conductivity} was obtained by converting the specific conductances to the equivalent conductances as follows. Since Ismail et al. prepared stock solution(s) of 0.997 mM solution at room temperature, the density polynomial proposed by Spedding et al.\cite{spedding1952conductances} was used to convert the concentration unit from molarity to molality at room temperature, and normality at different conditions were calculated based on the solvent densities calculated from Wagner-Pru\ss~equation.\cite{wagner2002iapws} The converted data show a reasonable agreement with our data; AAD at similar thermodynamic conditions is below 10 \%). 

\begin{figure}
	\includegraphics{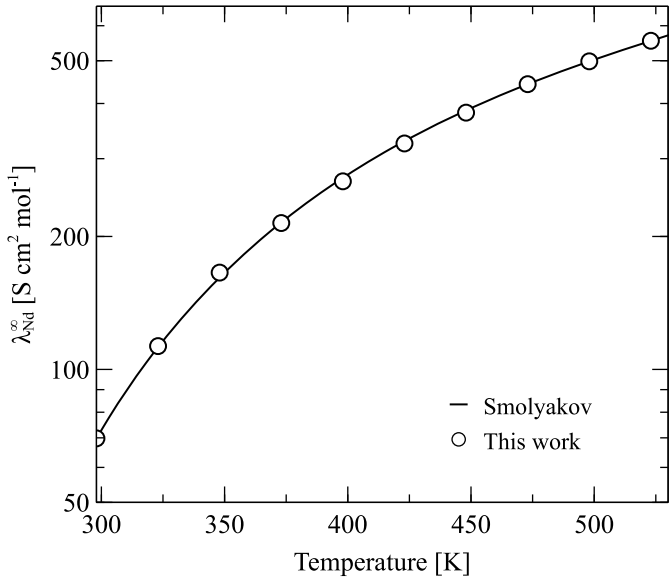}
	\caption{Comparison between the limiting conductances regressed from the TBBK model (open circles) and those from the Anderko-Lencka-Smolyakov equation (solid line). They agree well with each other.}
	\label{Fig5:limiting-conductance}
\end{figure}

Table \ref{tab2:limiting-conductance} shows the limiting conductance data obtained by fitting the TBBK equation. We compared the limiting conductance from the conductometric measurements to the Smolyakov-Anderko-Lencka corrrelation.\cite{anderko1997computation,smolyakov1975limiting}. It is given as:
\begin{equation}
	\ln\lambda_i^\infty(T)=A+B/T
	\label{eq:smolyakov}
\end{equation}
where $A$ and $B$ are adjustable parameters. Anderko and Lencka proposed a correlation between the structural entropy $\Delta S_\mathrm{str}^0$, a structural component of the hydration entropy,\cite{marcus1994viscosity} and the adjustable parameter $B$. Based on this correlation and the limiting conductance of $\mathrm{Nd^{3+}}$ at 298 K, the predictive equation [Eq. \ref{eq:smolyakov}] was used to calculate $\lambda_\mathrm{Nd}^\infty$. The limiting conductances experimentally estimated from our data are consistent with those from the Smolyakov-Anderko-Lencka correlation. (Figure \ref{Fig5:limiting-conductance})
\begin{table}
	\caption{Limiting conductances of the neodymium ion from 298 to 523 K along an isobar of 25 MPa.}
	\begin{tabular}{ccc}
		\hline
		$T$ [K] & $\lambda_\mathrm{Nd}^\infty$ (TBBK) [$\mathrm{S\ cm^{2}\ mol^{-1}}$] & $\lambda_\mathrm{Nd}^\infty$ (Eq. \ref{eq:smolyakov}) [$\mathrm{S\ cm^{2}\ mol^{-1}}$] \\ 
		\hline		
		298	&	69.8\textsuperscript{a}	&	70.0	\\
		323	&	112.9	$\pm$	3.0	&	112.6	\\
		348	&	165.7	$\pm$	2.6	&	161.8	\\
		373	&	214.5	$\pm$	1.1	&	215.5	\\
		398	&	266.6	$\pm$	1.7	&	271.9	\\
		423	&	324.8	$\pm$	2.3	&	329.2	\\
		448	&	381.4	$\pm$	3.4	&	386.3	\\
		473	&	442.4	$\pm$	7.7	&	442.7	\\
		498	&	498.4	$\pm$	16.2	&	498.7	\\
		523	&	554.7	$\pm$	20.3	&	555.0	\\
		\hline
		\multicolumn{3}{l}{\footnotesize\textsuperscript{a}0.1 MPa data from Apelblat\cite{apelblat2011representation} and Spedding et al.\cite{spedding1952conductances}}
	\end{tabular}
	\label{tab2:limiting-conductance}
\end{table}

Note that the Smolyakov equation is similar to the correlation proposed by Zimmerman, Arcis, and Tremaine (Eq. \ref{eq:ZAT-correlation}). When Eq. \ref{eq:ZAT-correlation} was fitted, the fitting parameters were obtained as $A_1=0.67$, $A_2=-0.97$ and $A_3=-13.97$ (AAD 0.69 \%).

Using the Smolyakov-Anderko-Lencka equation, the specific conductance data were generated based on different association constant models proposed by Haas et al.\cite{haas1995rare} and Migidsov and Williams-Jones\cite{migdisov2002spectrophotometric}.
\begin{figure}
	\includegraphics{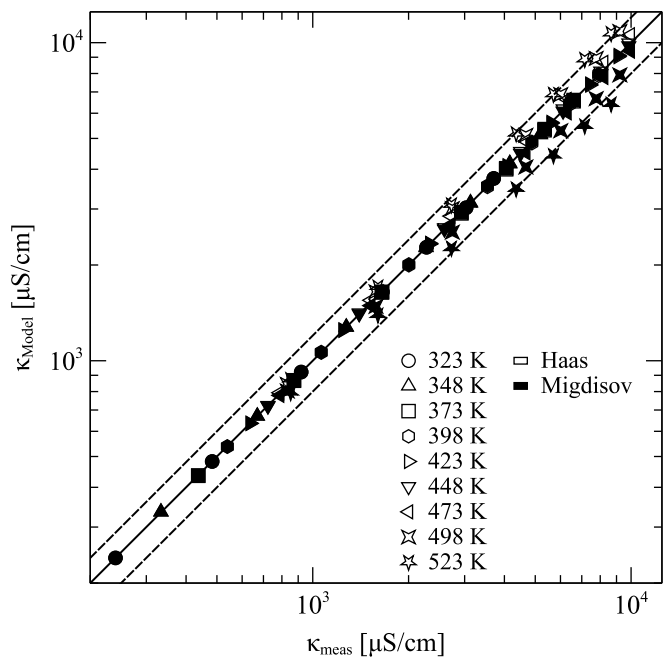}
	\caption{A parity plot between the specific conductance data obtained from this work and those calculated based on the association constants by Haas et al.\cite{haas1995rare} (open symbols) and Migdisov and Williams-Jones\cite{migdisov2002spectrophotometric} (filled symbols). At low temperatures, the empirical data and those calculated from two different models agree well with each other even when the association constants are different by a factor of $\mathcal{O}(10^2)$. At higher temeratures ($>398$ K), the measured conductance becomes higher than those calculated using the DEW parameters by Migdisov and Williams-Jones\cite{migdisov2002spectrophotometric} but lower than those from Haas et al.\cite{haas1995rare} The dashed lines represent AAD $\pm$15 \% lines. For numerical details, see the Table S3 in Supporting Information.}
	\label{Fig6:model-data-comparison}
\end{figure} 
Figure \ref{Fig6:model-data-comparison} compares the calculated specific conductance data to the measured ones. Although the second association constants are different by up to a factor of $\mathcal{O}(10^2)$, the specific conductance difference between these two models is comparable to the measurement uncertainties at low temperatures. This result suggests that the second association constants at low temperatures cannot be reliably derived. Indeed, when both the first and the second association steps are considered in model regression at low temperatures, the second association constant did not converge well or even became close to zero ($K_\mathrm{NdCl_2}\sim\mathcal{O}(10^{-7})$). 

\begin{figure}
	\includegraphics{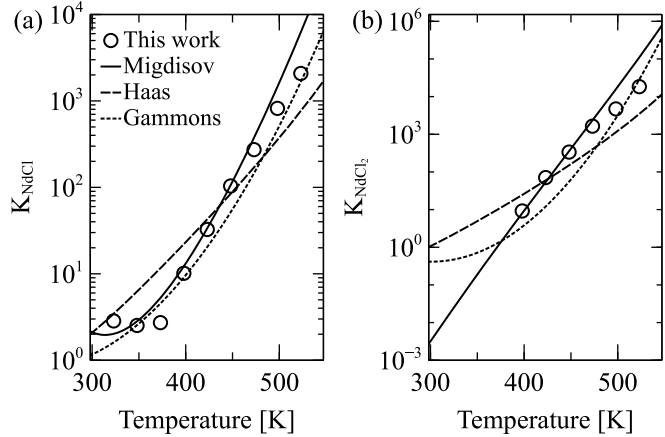}
	\caption{Comparison of association constants obtained from conductance measurements and those from earlier works. Both the first and the second association constants are close to those obtained by Migdisov and Williams-Jones\cite{migdisov2002spectrophotometric} and Gammons et al.\cite{gammons1996aqueous}}
	\label{Fig7:association-constants}
\end{figure}

Based on this observation, the equivalent conductance data obtained below 398 K were modeled with only the first association step $\mathrm{Nd^{3+}+Cl^{-}\rightleftharpoons NdCl^{2+}}$ (Table \ref{tab3:association-constants}). Figure \ref{Fig7:association-constants} compares the association constants calculated based on the DEW model\cite{migdisov2002spectrophotometric,haas1995rare,chan2021dewpython} and empirical correlations by Gammons et al.\cite{gammons1996aqueous}
\begin{table}
	\caption{Association constants obtained by regressing the TBBK model. At temperatures below 373 K, the second association constants were not calculated because they were statistically insignificant.}
	\begin{tabular}{ccc}
		\hline
		$T$ [K] & $K_\mathrm{NdCl}$ & $K_\mathrm{NdCl_{2}}$ \\
		\hline
		323	&	2.8	$\pm$	0.0	&	\textit{n/a}			\\
		348	&	2.5	$\pm$	0.1	&	\textit{n/a}			\\
		373	&	2.7	$\pm$	0.1	&	\textit{n/a}			\\
		398	&	10.1	$\pm$	0.5	&	9.2	$\pm$	0.5	\\
		423	&	32.6	$\pm$	4.9	&	70.8	$\pm$	22.0	\\
		448	&	103.8	$\pm$	5.3	&	336.5	$\pm$	156.3	\\
		473	&	273.1	$\pm$	21.6	&	1641.3	$\pm$	361.7	\\
		498	&	820.2	$\pm$	45.5	&	4705.7	$\pm$	820.7	\\
		523	&	2079.3	$\pm$	172.7	&	18358.8	$\pm$	1360.2	\\		
		\hline
	\end{tabular}
	\label{tab3:association-constants}
\end{table}

\begin{figure*}
	\includegraphics[width=\textwidth]{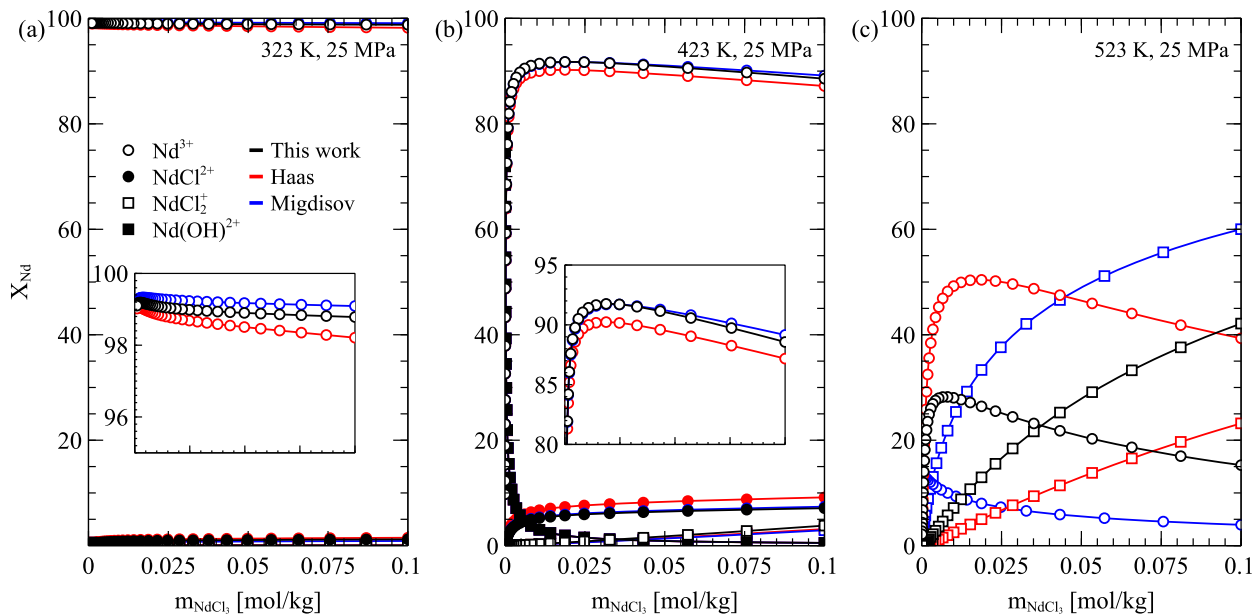}
	\caption{Ion speciation analysis based on different association constants at (a) 323 K, (b) 423 K and (c) 523 K along the 25 MPa isobar. Note that the fraction of free neodymium ions does not converge to 100 \%, since the complexation between neodymium and hydroxide ions cannot be ruled out. In all conditions, the concentration profiles from the association constants obtained in this work are closer to those obtained from Migdisov and Williams-Jones. In (a) and (b), insets magnify the difference in the fractions of free neodymium ions. In (c), the fractions of NdCl\textsuperscript{2+} and Nd(OH)\textsuperscript{2+} were not shown for visual clarity.}
	\label{Fig8:ion-speciation}
\end{figure*}

Ion speciation results can illustrate the ion pair formation process (Figure \ref{Fig8:ion-speciation}). At all temperatures, the fraction of the neodymium ions does \textit{not} converge to 100 \%, unlike other neutral cations (e.g., Na\textsuperscript{+}). This result arises because the ion-pair formation constants between neodymium and hydroxide ions are pretty high, even in room temperature solutions ($K_\mathrm{NdOH}\sim\mathcal{O}(10^6)$). The fraction of neodymium hydroxide complexes ($\mathrm{Nd(OH)}_n$), however, rapidly decreases as the concentration of neodymium chloride increases ($c_\mathrm{Cl}\gg c_\mathrm{OH}$). At the lowest temperature studied in this work (323 K), the fractions of neodymium ion species from different models are similar. At the highest temperature (523 K), the free ion fractions from different association constants are substantially different. The concentration profiles determined in this work are still closer to those calculated based on the DEW parameters provided by Migdisov and Williams-Jones. Hence, this work recommends using the DEW parameters proposed by Migdisov and Williams-Jones when modeling the behavior of neodymium chloride in aqueous solutions (instead of the original parameters by Haas et al.)

\section{Conclusions}
In this work, we used the in-situ conductometric device we designed in our earlier work\cite{yoon2020dielectric,yoon2021pyoecp,yoon2021situ} to measure complex impedance, from which the limiting conductances and the association constants were esetimated in dilute aqueous NdCl\textsubscript{3} solutions up to 523 K along an isobar of 25 MPa. The Turq-Blum-Bernard-Kunz model, which was modified based on several earlier works, was well fitted to the experimental conductance data. The limiting conductances obtained from the model regression agreed well with those calculated from the Smolyakov-Anderko-Lencka correlation, and the association constants were consistent with those obtained by Migdisov and Williams-Jones, and Gammons et al. These consistencies suggest that thermodynamic and kinetic properties (e.g., self-diffusion coefficient of ions in concentrated solutions\cite{dufreche2002ionic,bernard1992self}) can be readily calculated using the modified DEW model parameters and the Smolyakov-Anderko-Lencka correlation. Considering that these thermodynamic and kinetic properties are essential for process design, modeling, and simulations (e.g., in membrane distillation\cite{olatunji2018heat} and precipitation\cite{hodes2004salt}), experimental results and derived data reported, as well as the methodology presented in this work can serve as a guide for developing effective and environmentally-friendly REE recovery processes. 
\begin{acknowledgement}
This work was supported by the Director’s Postdoctoral Fellow Program
(20190653PRD4) and the Laboratory Directed Research and Development Program
(20190057DR) at Los Alamos National Laboratory.
\end{acknowledgement}

\begin{suppinfo}
Corrected expressions for the Turq-Blum-Bernard-Kunz (TBBK) model; Auxiliary equations for speciation analysis; Deep Earth Water (DEW) model parameters and empirical correlations for calculating association constants; and calculated conductance data based on different models.
\end{suppinfo}

\bibliography{bibliography}

\end{document}